# Tangential force and heating rate of a neutral relativistic particle mediated by equilibrium background radiation

G.V. Dedkov and A.A. Kyasov

*Nanoscale Physics Group, Kabardino –Balkarian State University, Nalchik, 360004, Russian Federation*

E-mail: gv_dedkov@mail.ru

A relativistic theory of fluctuation –electromagnetic interaction of a small neutral polarizable particle moving with respect to the equilibrium background radiation is developed. It is assumed that the particle radius is smaller then the characteristic wave length of the background radiation. General relativistic relations for the particle heating rate and tangential force are obtained and compared with the nonrelativistic and relativistic expressions of other authors.

## 1. Introduction

Interaction of a small neutral polarizable body with surrounding transparent medium filled by equilibrium background radiation (thermalized photon gas) is the well –known problem of fluctuation electrodynamics [1]. However, in the case of a particle moving with relativistic velocity $V \approx c$, correct description of the field structure both close to (at a distance of about or smaller than the radiation wave length and particle radius) and far from the particle (at a distance much larger than the particle radius) turns out to be not a trivial task. In this paper we consider the case when a particle moves relatively to the background radiation, whereas its radius is much smaller than characteristic Wien wave length of radiation, $R << \lambda_W = 2\pi \hbar c / k_B T$.

A simpler momentum transfer technique which has been used in earlier works to compute a net radiation force on a relativistic sphere moving through cosmic radiation [2], is not adequate in the case $R << \lambda_W$, because the particle interaction with photon gas can not be described in the framework of geometric optics which has been applied in [2].

The first attempt to calculate the drag force on a particle exerted by an equilibrium photon gas has been reported in [3] (see also [4]). However, the obtained result is valid only in the simplest case $V << c, T_1 = T_2 = T$, where $T_1$ and $T_2$ are the particle and background radiation temperature in their rest frames. General solution to this problem has been obtained in our paper [5] using the relativistic fluctuation electrodynamics. The developed formalism has been also used to calculate fluctuation electromagnetic forces and heating rate of a relativistic particle moving close to a flat surface of homogeneous and isotropic polarizable medium [6-8].

Quite recently, the corresponding problem has been considered in [9], where the authors have obtained the results essentially different from our's [5-7]. The aim of this work is to develop our theory [5] in more details and to feature principal drawbacks of approach used in [9].

## 2. Theory

Consider a spherical particle of radius $R$ having dipole electric and magnetic polarizabilities $\alpha_{e,m}(\omega)$ and temperature $T_1$, moving in $x-$ direction of the Cartesian coordinate system related to the stationary background radiation of temperature $T_2$ (fig.1) The particle may be considered as a point –like fluctuating dipole at

$$\frac{k_B R}{2\pi \hbar c} \max(T_1, T_2) << 1, \tag{1}$$

where $k_B$ and $\hbar$ are Boltzmann's and Planck's constants. Following [5-8], the fluctuation electromagnetic tangential force $F_x$ and rate of particle heating (cooling) $\dot{Q}$ are given by (both determined in the frame $\Sigma$ of the resting background):

$$F_x = \left\langle \nabla_x \left( \mathbf{d}^{sp} \cdot \mathbf{E}^{ind} + \mathbf{m}^{sp} \cdot \mathbf{B}^{ind} \right) \right\rangle + \left\langle \nabla_x \left( \mathbf{d}^{ind} \cdot \mathbf{E}^{sp} + \mathbf{m}^{ind} \cdot \mathbf{B}^{sp} \right) \right\rangle \equiv F_x^{(1)} + F_x^{(2)}, \tag{2}$$

$$\dot{Q} = \left\langle \left( \dot{\mathbf{d}}^{sp} \cdot \mathbf{E}^{ind} + \dot{\mathbf{m}}^{sp} \cdot \mathbf{B}^{ind} \right) \right\rangle + \left\langle \left( \dot{\mathbf{d}}^{ind} \cdot \mathbf{E}^{sp} + \dot{\mathbf{m}}^{ind} \cdot \mathbf{B}^{sp} \right) \right\rangle \equiv \dot{Q}_1 + \dot{Q}_2, \tag{3}$$

where the superscripts "sp", "ind" denote spontaneous and induced components of the fluctuating dipole electric and magnetic moments **d, m,** and electromagnetic fields **E, B**, points above $Q, \mathbf{d}, \mathbf{m}$ denote time differentiation, and the angular brackets denote complete quantum and statistical averaging. All vectorial quantities in (2), (3) are assumed to be the Heisenberg operators corresponding to the reference frame $\Sigma$. The electric $\mathbf{P}(\mathbf{r},t)$ and magnetic $\mathbf{M}(\mathbf{r},t)$ polarization vectors produced by a moving particle are given by

$$\mathbf{P}(\mathbf{r},t) = \mathbf{d}(t)\delta(\mathbf{r} - \mathbf{V}t) \tag{4}$$

$$\mathbf{M}(\mathbf{r},t) = \mathbf{m}(t)\delta(\mathbf{r} - \mathbf{V}t) \tag{5}$$

Passing in Eqs. (2) and (3) to the Fourier transforms for all spatial coordinates and time, we write the Maxwell equations for the Fourier transforms of the Hertz vectors of electromagnetic field induced by electric and magnetic dipole moments of the particle ($\Delta = -k_x^2 - k_y^2 - k_z^2 = -k^2$):

$$\left(\Delta + \frac{\omega^2}{c^2}\varepsilon(\omega)\right)\mathbf{\Pi}^e_{\omega\mathbf{k}} = -\frac{4\pi}{\varepsilon(\omega)}\mathbf{P}_{\omega\mathbf{k}} \qquad (6)$$

$$\left(\Delta + \frac{\omega^2}{c^2}\varepsilon(\omega)\right)\mathbf{\Pi}^m_{\omega\mathbf{k}} = -4\pi\mathbf{M}_{\omega\mathbf{k}} \qquad (7)$$

For vacuum, we have $\mu(\omega) = 1$ and $\varepsilon(\omega) = 1 + i\cdot\eta\cdot sign(\omega)$, assuming the limit $\eta \to 0$ to be applied in the final results [1]. The Fourier transforms of the quantities in Eqs. (6),(7) are defined in the reference frame $\Sigma$ (see fig. 1)

$$\mathbf{X}_{\omega\mathbf{k}} = \iint d^3r\, dt\, \mathbf{X}(\mathbf{r},t)\exp[-i(\mathbf{k}\mathbf{r} - \omega t)] \qquad (8)$$

According to Eqs. (4),(5), in order to obtain the Fourier transforms $\mathbf{P}_{\omega\mathbf{k}}$ and $\mathbf{M}_{\omega\mathbf{k}}$ of the polarization and magnetization vectors, we should express the spontaneous moments $\mathbf{d}^{sp}(t), \mathbf{m}^{sp}(t)$ of the particle in the reference frame $\Sigma$. For this purpose, we use the well-known relativistic transformations of the corresponding quantities from the particle rest frame $\Sigma'$ to the frame $\Sigma$ (see Eqs.(A7), (A8) in Appendix A)

Writing the Fourier expansions for $d_x^{sp'}(t')$ and $m_x^{sp'}(t')$ in the reference frame $\Sigma'$ of the particle, where the frequency $\omega'$ and time $t'$ are expressed in terms of $\omega$ and $t$ in the reference frame $\Sigma$, we then substitute them into Eqs. (A7) and (A8). It is easy to verify, for example, that an $x-$ projection of the spontaneous dipole moment $d_x^{sp}(t)$ is given by

$$d_x^{sp}(t) = (2\pi)^{-1}\int_{-\infty}^{\infty}d\omega\, d_x^{sp'}(\gamma\omega)\exp(-i\omega t) \qquad (9)$$

The integral representations for all projections of spontaneous electric and magnetic dipole moments are listed in Appendix A.

With account of the obtained spontaneous dipole moments being substituted into Eqs. (4) and (5), we find the Fourier transforms $\mathbf{P}_{\omega\mathbf{k}}$ and $\mathbf{M}_{\omega\mathbf{k}}$ according to Eq. (8). Furthermore, the Fourier transforms of the respective Hertz vectors of electromagnetic field are found from Eqs. (6) and (7). The final results obtained after passing to the limiting case of vacuum are presented in Appendix B.

Next, we find out the Fourier transforms of the induced electric and magnetic fields associated with a moving particle. For this purpose, we use the results of Appendix B and the known relations [10]

$$\mathbf{E}^{ind}_{\omega \mathbf{k}} = rot\, rot\, \mathbf{\Pi}^{e}_{\omega \mathbf{k}} + \frac{i\omega}{c} rot\, \mathbf{\Pi}^{m}_{\omega \mathbf{k}} \qquad (10)$$

$$\mathbf{H}^{ind}_{\omega \mathbf{k}} = rot\, rot\, \mathbf{\Pi}^{m}_{\omega \mathbf{k}} - \frac{i\omega}{c} \varepsilon(\omega) rot\, \mathbf{\Pi}^{e}_{\omega \mathbf{k}} \qquad (11)$$

The Fourier integrals for induced fields and spontaneous moments of the particle in the reference frame $\Sigma$ should be substituted into Eq. (2),

$$F_x^{(1)} = \left\langle \nabla_x \left( \mathbf{d}^{sp} \cdot \mathbf{E}^{ind} + \mathbf{m}^{sp} \cdot \mathbf{H}^{ind} \right) \right\rangle \qquad (12)$$

The correlators of dipole moments arising under statistical averaging can be found from the fluctuation dissipation relations being written in the particle rest frame [11]

$$\left\langle d_i^{sp'}(\omega) d_k^{sp'}(\omega') \right\rangle = 2\pi \delta_{ik} \delta(\omega+\omega') \hbar \alpha_e''(\omega) \coth\frac{\hbar \omega}{2k_B T_1} \qquad (13)$$

$$\left\langle m_i^{sp'}(\omega) m_k^{sp'}(\omega') \right\rangle = 2\pi \delta_{ik} \delta(\omega+\omega') \hbar \alpha_m''(\omega) \coth\frac{\hbar \omega}{2k_B T_1} \qquad (14)$$

where $\alpha_{e,m}''(\omega)$ are the imaginary parts of the electric and magnetic polarizabilities, respectively. It should be noted that coordinates of a moving particle ($Vt$, 0, 0) have to be substituted into Eq. (12) only after differentiation with respect to $x$. To circumvent singularities associated with the resonant denominators in expressions for the Fourier transforms of the Hertz vectors (Appendix B), the following relation is used

$$\left(k^2 - \omega^2/c^2 - i\cdot 0\cdot sign\omega\right)^{-1} = P\left[\left(k^2-\omega^2/c^2\right)^{-1}\right] + i\cdot \pi\, \delta\!\left(k^2-\omega^2/c^2\right) sign\omega =$$
$$= P\left[\left(k^2-\omega^2/c^2\right)^{-1}\right] - \frac{i\pi}{2k}\left[\delta(\omega/c+k) - \delta(\omega/c-k)\right] \qquad (15)$$

where $P(...)$ is the principal value of the corresponding integral. The procedure described above leads to the following result ($\beta = V/c$):

$$F_x^{(1)} = \frac{\gamma \hbar}{\pi c^4} \int_0^\infty d\omega\, \omega^4 \int_{-1}^1 dx\, x(1+\beta x)^2 \coth\left[\frac{\gamma \hbar \omega(1+\beta x)}{2k_B T_1}\right] \cdot \left[\alpha_e''(\gamma\omega(1+\beta x)) + \alpha_m''(\gamma\omega(1+\beta x))\right] \qquad (16)$$

A contribution from induced dipole moments of the particle (or, equivalently, from spontaneous fluctuating fields $\mathbf{E}^{sp}$ and $\mathbf{H}^{sp}$) to the tangential force is given by the second part of Eq. (2),

$$F_x^{(2)} = \left\langle \nabla_x \left( \mathbf{d}^{ind} \cdot \mathbf{E}^{sp} + \mathbf{m}^{ind} \cdot \mathbf{H}^{sp} \right) \right\rangle \qquad (17)$$

In order to perform statistical averaging in (17), it is necessary to calculate the induced electric and magnetic moments of the particle. For this purpose, we use integral relations describing a temporal dispersion between spontaneous fields of the background and induced moments of the particle in the particle rest frame [12]

$$\mathbf{d}^{ind\,\prime}(t') = \int_{-\infty}^{t'} \alpha_e(t'-\tau') \mathbf{E}^{sp\,\prime}(\mathbf{r}';\tau')d\tau' \qquad (18)$$

$$\mathbf{m}^{ind\,\prime}(t') = \int_{-\infty}^{t'} \alpha_m(t'-\tau') \mathbf{H}^{sp\,\prime}(\mathbf{r}';\tau')d\tau' \qquad (19)$$

Applying relativistic transformations for the electric and magnetic fields and substituting Eqs.(18),(19) into Eqs.(A7),(A8), we obtain the integral relations between the induced moments of the particle and the Fourier transforms of the fields in the reference frame $\Sigma$. For example, using (A7) and (18) and relativistic transformation $E'_x = E_x$, we find the induced moment $d_x^{ind}(t)$ to be

$$d_x^{ind}(t) = \gamma^{-1}(2\pi)^{-4} \iint d\omega\, d^3k\, \alpha_e[\gamma(\omega - k_x V)] E^{sp}_{x,\omega\mathbf{k}} \cdot \exp[-i(\omega - k_x V)t] \qquad (20)$$

The expressions for all projections of the induced electric and magnetic moments are presented in Appendix C.

The Fourier transforms of the operators $\mathbf{d}^{ind}(t), \mathbf{m}^{ind}(t)$ and vacuum fields are substituted into Eq. (17). The arising correlators of the Fourier transforms of electric and magnetic fields involved can be expressed in terms of the corresponding spectral densities using the relations following from the condition of stationarity of electromagnetic fluctuations

$$\left\langle U_i^{sp}{}_{\omega\mathbf{k}} V_j^{sp}{}_{\omega'\mathbf{k}'} \right\rangle = (2\pi)^4 \delta(\omega+\omega')\delta(\mathbf{k}+\mathbf{k}') \left( U_i^{sp} V_j^{sp} \right)_{\omega\mathbf{k}} \qquad (21)$$

where $U_i^{sp}, V_i^{sp} = E_i^{sp}, B_i^{sp}$ (i, j = x, y, z).

The spectral densities of the fields in (21) have to be expressed through the retarded Green's function for a homogeneous isotropic nonmagnetic medium [1]

$$\left(E_i^{sp} E_k^{sp}\right)_{\omega \mathbf{k}} = -\coth\frac{\hbar\omega}{2k_B T}\frac{\omega^2}{c^2}\operatorname{Im} D_{ik}(\omega,\mathbf{k}) \qquad (22)$$

$$\left(B_i^{sp} B_k^{sp}\right)_{\omega \mathbf{k}} = -\coth\frac{\hbar\omega}{2k_B T} rot_{il} rot'_{km} \operatorname{Im} D_{lm}(\omega,\mathbf{k}) \qquad (23)$$

$$\left(E_i^{sp} B_k^{sp}\right)_{\omega \mathbf{k}} = -\coth\frac{\hbar\omega}{2k_B T}\frac{i\omega}{c} rot'_{km} \operatorname{Im} D_{lm}(\omega,\mathbf{k}) \qquad (24)$$

In Eqs.(22)-(24), $D_{ik}(\omega,\mathbf{k})$ denotes the Fourier transform of the retarded Green's function of vacuum [1]

$$D_{ik}^{R}(\omega,\mathbf{k}) = \frac{4\pi\hbar}{\omega^2/c^2 - k^2 + i\cdot 0\cdot sign(\omega)}\left[\delta_{ik} - \frac{c^2 k_i k_k}{\omega^2}\right] \qquad (25)$$

An imaginary part of $D^{R}_{ik}(\omega,\mathbf{k})$ is related to the rule for the round of the poles $\omega = \pm ck$, Eq. (15). Using Eqs. (21)-(24) and Appendix C, Eq.(17) takes the form

$$F_x^{(2)} = -\frac{\gamma\hbar}{\pi c^4}\int_0^\infty d\omega\,\omega^4 \int_{-1}^{1} dx\,x(1+\beta x)^2 \coth\left[\frac{\hbar\omega}{2k_B T_2}\right]\cdot\left[\alpha_e''(\gamma\omega(1+\beta x)) + \alpha_m''(\gamma\omega(1+\beta x))\right] \qquad (26)$$

Finally, summing Eqs.(16) and (26), we obtain

$$F_x = -\frac{\gamma\hbar}{\pi c^4}\int_0^\infty d\omega\,\omega^4 \int_{-1}^{1} dx\,x(1+\beta x)^2 \cdot \left[\alpha_e''(\gamma\omega(1+\beta x)) + \alpha_m''(\gamma\omega(1+\beta x))\right]\cdot$$
$$\cdot\left[\coth\left(\frac{\hbar\omega}{2k_B T_2}\right) - \coth\left(\frac{\gamma\hbar\omega(1+\beta x)}{2k_B T_1}\right)\right] \qquad (27)$$

Eq.(27) has been firstly presented in [5] without magnetic polarization term. It should be noted that each of Eqs.(16) and (26) contains divergent integrals over frequency and only their sum gives a physically correct result, Eq.(27). In a linear velocity approximation, Eq.(27) reduces to

$$F_x = -\frac{4\hbar V}{3\pi c^5} \int_0^\infty d\omega\, \omega^5 \cdot \left\{ \begin{array}{l} \frac{\hbar}{4k_B T_1} \frac{\alpha_e''(\omega) + \alpha_m''(\omega)}{\sinh^2(\hbar\omega/2k_B T_1)} + 2[\Pi(\omega,T_2) - \Pi(\omega,T_1)] \cdot \\ \cdot\left[\frac{\alpha_e''(\omega) + \alpha_m''(\omega)}{\omega} + \frac{1}{2}\frac{d\alpha_e''(\omega)}{d\omega} + \frac{1}{2}\frac{d\alpha_m''(\omega)}{d\omega}\right] \end{array} \right\} \quad (28)$$

Particularly, at $T_1 = T_2 = T$, Eq.(28) simplifies to

$$F_x = -\frac{\hbar^2}{3\pi c^4}\frac{\beta}{k_B T}\int_0^\infty d\omega\, \omega^5 [\alpha_e''(\omega) + \alpha_m''(\omega)]\sinh^{-2}(\hbar\omega/2k_B T) \quad (29)$$

Eq.(29) without the magnetic polarization term coincides with the result [3] obtained using a nonrelativistic approach. However, it is worth noting that the authors of [3] completely neglected the contribution to the interaction coming from the spontaneous dipole moment of a particle. A finite result was obtained only because the calculation was carried out in the rest frame of the particle. We will discuss this nontrivial issue in Section 3.

The heating rate of the particle is calculated with the help of Eq. (3), quite analogous to the calculation of the tangential force. The final result is given by [5-7]

$$\dot{Q} = \langle \dot{\mathbf{d}}^{sp}\mathbf{E}^{ind} + \dot{\mathbf{m}}^{sp}\mathbf{H}^{ind}\rangle + \langle \dot{\mathbf{d}}^{ind}\mathbf{E}^{sp} + \dot{\mathbf{m}}^{ind}\mathbf{H}^{sp}\rangle = \dot{Q}_1 + \dot{Q}_2 =$$
$$= \frac{\gamma\hbar}{\pi c^3}\int_0^\infty d\omega\,\omega^4 \int_{-1}^1 dx\, x(1+\beta x)^3 \cdot [\alpha_e''(\gamma\omega(1+\beta x)) + \alpha_m''(\gamma\omega(1+\beta x))]\cdot \quad (30)$$
$$\cdot\left[\coth\left(\frac{\hbar\omega}{2k_B T_2}\right) - \coth\left(\frac{\gamma\hbar\omega(1+\beta x)}{2k_B T_1}\right)\right]$$

As in Eqs. (16) and (26), the terms in (30) related with $T_1$, describe the contributions of spontaneous dipole moments of the particle, while the terms related with $T_2$ describe the contributions of induced dipole moments of the particle or, equivalently –spontaneous fluctuations of the background electromagnetic field. At $\beta = 0, \gamma = 1$, the simplest form of Eq.(30) reads

$$\dot{Q} = \frac{2\hbar}{\pi c^3}\int_0^\omega d\omega\,\omega^4 \cdot [\alpha_e''(\omega) + \alpha_m''(\omega)] \cdot \left[\coth\left(\frac{\hbar\omega}{2k_B T_2}\right) - \coth\left(\frac{\hbar\omega}{2k_B T_1}\right)\right] \quad (31)$$

Eq.(31) can also be derived in a very simple way using the energy conservation and Kirchhoff's laws. The heating rate $\dot{Q}$ can be cast in the form

$$\dot{Q} = I(T_2) - I(T_1) \quad , \quad (32)$$

where $I(T_1)$ is given by

$$I(T_1) = \frac{2}{3c^3}\left(\left\langle(\ddot{\mathbf{d}}^{sp}(t))^2\right\rangle + \left\langle(\ddot{\mathbf{m}}^{sp}(t))^2\right\rangle\right) \tag{33}$$

Eq.(33) describes an average intensity of the dipole radiation of the particle (with temperature $T_1$) in vacuum space caused by spontaneous fluctuating moments. Evidently, this dipolar radiation leads to the particle cooling, while the function $I(T_1)$ is determined by the particle absorption spectrum and temperature $T_1$. On the other hand, $I(T_2)$ represents an average intensity of the vacuum radiation, illuminating the particle. According to the Kirchhoff's law, the function $I(T_2)$ should have the same functional form as the function $I(T_1)$. The only difference between both functions is related with different temperatures. This allows to calculate only the function $I(T_1)$, whereas the function $I(T_2)$ can be obtained replacing $T_1$ by $T_2$ in the expression for $I(T_1)$. Thus, for the dipole electric moment we get

$$\left\langle(\ddot{\mathbf{d}}^{sp}(t))^2\right\rangle = \int_{-\infty}^{\infty}\frac{d\omega}{2\pi}\int_{-\infty}^{\infty}\frac{d\omega'}{2\pi}\left\langle\ddot{\mathbf{d}}^{sp}(\omega)\ddot{\mathbf{d}}^{sp}(\omega')\right\rangle\exp\left[-i(\omega+\omega')t\right], \tag{34}$$

Then, bearing in mind an identity of the reference frames $\Sigma$ and $\Sigma'$, with account of (13) we get

$$\left\langle(\ddot{\mathbf{d}}^{sp}(t))^2\right\rangle = \frac{3\hbar}{\pi}\int_0^{\infty}d\omega\,\omega^4\alpha_e''(\omega)\coth(\hbar\omega/2k_B T_1) \tag{35}$$

In the same manner, using Eq.(14), the contribution from spontaneous magnetic moment is given by

$$\left\langle(\ddot{\mathbf{m}}^{sp}(t))^2\right\rangle = \frac{3\hbar}{\pi}\int_0^{\infty}d\omega\,\omega^4\alpha_m''(\omega)\coth(\hbar\omega/2k_B T_1) \tag{36}$$

From (33)- (36) we finally get

$$I(T_1) = \frac{2\hbar}{\pi c^3}\int_0^{\infty}d\omega\,\omega^4(\alpha_e''(\omega)+\alpha_m''(\omega))\coth(\hbar\omega/2k_B T_1) \tag{37}$$

The function $I(T_2)$, according to the Kirchhoff's law, reads

$$I(T_2) = \frac{2\hbar}{\pi c^3} \int_0^\infty d\omega \omega^4 (\alpha_e''(\omega) + \alpha_m''(\omega)) \coth(\hbar\omega/2k_B T_2) \tag{38}$$

Introducing Eqs.(37) and (38) into (32), we explicitly retrieve Eq. (31). To our knowledge, despite quite obvious character of Eq.(31), even this simple result has been unknown prior to our paper [5]. The physical essence of the above consideration allows to understand principal importance of spontaneous and induced fluctuation moments of the particle in the problem of fluctuation electromagnetic interaction. This general conclusion concerns both calculation of the heating rate $\dot{Q}$ and the force $F_x$.

## 3. Discussion

Now let us discuss a non trivial fact that Eq. (29) coincides with Eq.(27) at $V \ll c, T_1 = T_2 = T$. Remember that Eq.(27) has been obtained from the general relativistic theory, with total account of the spontaneous and induced moments of the particle, whereas Eq.(29) can be derived with neglect of spontaneous fluctuations of the particle [3], simply performing a calculation in the rest frame of the particle, $\Sigma'$ (see fig.1). This proves to be true only in the case $T_1 = T_2 = T$, and the reason is as follows. It turns out that $F_x'(T_1)$ becomes finite quantity in the reference frame $\Sigma'$ (despite being divergent in the reference frame $\Sigma$), whereas at $V \ll c$, $T_1 = T_2 = T$, $F_x'(T_1) = 0$. Therefore, the second contribution to the tangential force, $F_x'(T_2)$, turns out to be unique in the reference frame $\Sigma'$. An identity of the resulting expressions for the tangential force in both reference frames $\Sigma$ and $\Sigma'$ stems from the relativistic transformation law for the tangential force $F_x$:

$$F_x = \frac{F_x' + (\beta/c)(\mathbf{F}'\mathbf{V}')}{1 + (\beta/c)V_x'} \tag{39}$$

Because $\mathbf{V}' = 0$ in $\Sigma'$, we get from (39) $F_x = F_x'$. Therefore, the force component $F_x'(T_2)$ being calculated in $\Sigma'$, equals the total dissipative force in $\Sigma$, as well. In the general case, at $T_1 \neq T_2$ and relativistic velocity of the particle, this conclusion is invalid, and the relation $F_x = F_x'$ can be applied only with respect to the total force $F_x = F_x'(T_1) + F_x'(T_2)$, which must be calculated with account of both spontaneous and induced fluctuating dipole moments. Therefore, it is quite obvious that a relativistic generalization of the method [3], with no account of the spontaneous fluctuation moments of the particle $\mathbf{d}^{sp}$, $\mathbf{m}^{sp}$, can not lead to correct expression for the tangential force in the reference frame $\Sigma$.

Just this error has been done in the recent paper by Volokitin and Persson [9], where they first computed $F_x(T_2) = \langle \nabla_x (\mathbf{d}^{ind} \mathbf{E}^{sp} + \mathbf{m}^{ind} \mathbf{H}^{sp}) \rangle$ in $\Sigma'$, and then transformed this expression to the reference frame $\Sigma$, using the relationship $F_x(T_2) = F'_x(T_2)$. As a result, these authors have obtained the vacuum tangential force to be given by Eq.(49) in ref.[9]. Writing this one in our notations yields

$$F'_x = \frac{\hbar}{\pi c^4} \int_0^\infty d\omega \omega^4 \int_{-1}^1 dx x (\alpha''_e(\omega) + \alpha''_m(\omega)) \coth\left(\frac{\hbar \gamma \omega(1+\beta x)}{2 k_B T_2}\right) \qquad (40)$$

One sees that Eq.(40) does not depend on the particle temperature $T_1$, i.e. it is independent of spontaneous fluctuations of the particle. Moreover, comparing Eq.(40) with (27) reveals other differences : the lack of factors $\gamma$ and $(1+\beta x)^2$, the presence of frequency $\omega$ instead $\gamma\omega(1+\beta x)$ in the arguments of $\alpha''_{e,m}(\omega)$, and incorrect Doppler –shifted frequency under the sign of hyperbolic cotangent. In this case, the Doppler –shifted frequency can not appear because the photon gas is assumed to be in rest in the reference frame $\Sigma$, by definition. All these drawbacks are due to complete neglect of spontaneous dipole moments of the particle and incorrect using of the relation $F'_x(T_2) = F'_x(T_2)$, because generally speaking, $F_x = F'_x$ only for the total tangential force : $F_x = F_x(T_1) + F_x(T_2)$.

Furthermore, if use is made of the same method to calculate the heating rate $\dot{Q}$, then instead of Eq.(31) (i.e. at $V = 0, \gamma = 1$) we get the divergent expression

$$\dot{Q} = \frac{2\hbar}{\pi c^3} \int_0^\omega d\omega \omega^4 [\alpha''_e(\omega) + \alpha''_m(\omega)] \coth\left(\frac{\hbar \omega}{2 k_B T_2}\right) \to \infty \qquad (41)$$

Even if, formally, to exclude from Eq.(41) a contribution from zero-point modes, replacing $\coth\left(\frac{\hbar\omega}{2k_B T_2}\right)$ by $2(\exp(\hbar\omega/k_B T_2) - 1)^{-1}$, then Eq.(41) conflicts with the first and second laws of thermodynamics: $\dot{Q} > 0$ irrespectively of the relation between $T_1$ and $T_2$, $\int \dot{Q} dt \to \infty, T_1 \to \infty$. Therefore, an attempt of relativistic generalization of the result [3] made in [9], results in catastrophic consequences.

It is worth noting that Eqs.(27) and (28) predict a possibility for the particle acceleration [5], but, contrary to the claim of the authors [9] on this occasion, the corresponding acceleration effect is not caused by particle thermal radiation (being isotropic in the reference frame $\Sigma'$). That is because the heating rate $\dot{Q}$, the power of tangential force $\mathbf{F}\cdot\mathbf{V}$ and rate of energy change of the fluctuating electromagnetic field prove to be related by the energy conservation law, assuming whole system to be stationary [5-7]:

$$-\frac{dW}{dt} = \frac{dQ}{dt} + \mathbf{F}\cdot\mathbf{V} \qquad (42)$$

where $W$ is the energy of the fluctuating electromagnetic field. The quantities entering Eq.(42) have to be calculated in a joint manner. Particularly, a possibility of simultaneous acceleration and heating of the particle is provided by the work performed by electromagnetic field, resulting in the total energy decrease of the field. Other situations are also possible.

An important problem seems to obtain a self consistent solution of the relativistic dynamics equation $mcd\beta/dt = \gamma^{-3/2}F_x(\beta,T_1)$ and equation $CdT_1/dt' = \gamma^2 \dot{Q}(\beta,T_1)$ for the temperature evolution of the particle in its rest frame, where $C$ is the specific heat of the particle material. Several limiting cases have been considered in [5], whereas a more detailed analysis is still in progress.

### 4.Conclusion

A relativistic theory of the fluctuation electromagnetic interaction of a small neutral polarizable particle moving through an equilibrium background radiation (photon gas) is developed. Using the dipole approximation, the expressions for tangential force and heating rate of the particle are given. We draw a special attention to the fact that correct formulas for these quantities must incorporate both effects of induced and spontaneous fluctuating moments of the particle. The obtained results take into account arbitrary velocity, electric and magnetic polarization of the particle depending on material properties, and different temperatures of the particle and background. The developed theory predicts a possibility for the particle acceleration at definite conditions. Moreover, we show that recently obtained relativistic expression [9] for the drag force on a particle moving through the background radiation is in error.

**Appendix A**

**Projections of the spontaneous electric and magnetic moments of a particle in the reference frame $\Sigma$**

$$d_x^{sp}(t) = (2\pi)^{-1} \int_{-\infty}^{\infty} d\omega\, d_x^{sp'}(\gamma\omega) \exp(-i\omega t) \quad (A1)$$

$$d_y^{sp}(t) = (2\pi)^{-1} \gamma \int_{-\infty}^{\infty} d\omega \left[ d_y^{sp'}(\gamma\omega) - \beta m_z^{sp'}(\gamma\omega) \right] \exp(-i\omega t) \quad (A2)$$

$$d_z^{sp}(t) = (2\pi)^{-1} \gamma \int_{-\infty}^{\infty} d\omega \left[ d_z^{sp'}(\gamma\omega) + \beta m_y^{sp'}(\gamma\omega) \right] \exp(-i\omega t) \quad (A3)$$

$$m_x^{sp}(t) = (2\pi)^{-1} \int_{-\infty}^{\infty} d\omega\, m_x^{sp'}(\gamma\omega) \exp(-i\omega t) \quad (A4)$$

$$m_y^{sp}(t) = (2\pi)^{-1}\gamma \int_{-\infty}^{\infty} d\omega \left[ m_y^{sp'}(\gamma\omega) + \beta d_z^{sp'}(\gamma\omega) \right] \exp(-i\omega t) \tag{A5}$$

$$m_z^{sp}(t) = (2\pi)^{-1}\gamma \int_{-\infty}^{\infty} d\omega \left[ m_z^{sp'}(\gamma\omega) - \beta d_y^{sp'}(\gamma\omega) \right] \exp(-i\omega t) \tag{A6}$$

In order to get projections of the dipole moments (A1)-(A6), we used the known relatibistic transformations

$$\mathbf{d} = \mathbf{d}' + \frac{1}{c}[\mathbf{V}\,\mathbf{m}'] - \frac{(\gamma-1)}{\gamma}\frac{\mathbf{V}(\mathbf{V}\cdot\mathbf{d}')}{V^2} \tag{A7}$$

$$\mathbf{m} = \mathbf{m}' - \frac{1}{c}[\mathbf{V}\,\mathbf{d}'] - \frac{(\gamma-1)}{\gamma}\frac{\mathbf{V}(\mathbf{V}\cdot\mathbf{m}')}{V^2} \tag{A8}$$

**Appendix B**
**Fourier transforms of the Hertz vectors of the induced electromagnetic field in the reference frame $\Sigma$**

$$\Pi^e_{x,\omega\mathbf{k}} = 4\pi \frac{d_x^{sp'}(\gamma\omega^-)}{k^2 - \omega^2/c^2 - i\cdot 0\cdot \mathrm{sign}\,\omega} \tag{B1}$$

$$\Pi^e_{y,\omega\mathbf{k}} = 4\pi\gamma \frac{d_y^{sp'}(\gamma\omega^-) - \beta m_z^{sp'}(\gamma\omega^-)}{k^2 - \omega^2/c^2 - i\cdot 0\cdot \mathrm{sign}\,\omega} \tag{B2}$$

$$\Pi^e_{z,\omega\mathbf{k}} = 4\pi\gamma \frac{d_z^{sp'}(\gamma\omega^-) + \beta m_y^{sp'}(\gamma\omega^-)}{k^2 - \omega^2/c^2 - i\cdot 0\cdot \mathrm{sign}\,\omega} \tag{B3}$$

$$\Pi^m_{x,\omega\mathbf{k}} = 4\pi \frac{m_x^{sp'}(\gamma\omega^-)}{k^2 - \omega^2/c^2 - i\cdot 0\cdot \mathrm{sign}\,\omega} \tag{B4}$$

$$\Pi^m_{y,\omega\mathbf{k}} = 4\pi\gamma \frac{m_y^{sp'}(\gamma\omega^-) + \beta d_z^{sp'}(\gamma\omega^-)}{k^2 - \omega^2/c^2 - i\cdot 0\cdot \mathrm{sign}\,\omega} \tag{B5}$$

$$\Pi^m_{z,\omega\mathbf{k}} = 4\pi\gamma \frac{m_z^{sp'}(\gamma\omega^-) - \beta d_y^{sp'}(\gamma\omega^-)}{k^2 - \omega^2/c^2 - i\cdot 0\cdot \mathrm{sign}\,\omega} \tag{B6}$$

where $\omega^- = \omega - k_x V$. Note that projections of the Fourier components of the dipole moments $d'_{x,y,z}$ and $m'_{x,y,z}$ are taken in the particle reference frame $\Sigma'$.

**Appendix C**
**Induced electric and magnetic moments of a particle in the reference frame $\Sigma$**

$$d_x^{ind}(t) = \gamma^{-1}(2\pi)^{-4} \iint d\omega\, d^3k\, \alpha_e(\gamma\omega^-) E^{sp}_{x,\omega\mathbf{k}} \cdot \exp(-i\omega^- t) \tag{C1}$$

$$d_y^{ind}(t) = \gamma(2\pi)^{-4} \iint d\omega\, d^3k \left\{ \begin{array}{l} \alpha_e(\gamma\omega^-)[E^{sp}_{y,\omega\mathbf{k}} - \beta B^{sp}_{z,\omega\mathbf{k}}] - \\ -\beta\alpha_m(\gamma\omega^-)[B^{sp}_{z,\omega\mathbf{k}} - \beta E^{sp}_{y,\omega\mathbf{k}}] \end{array} \right\} \cdot \exp(-i\omega^- t) \tag{C2}$$

$$d_z^{ind}(t) = \gamma(2\pi)^{-4} \iint d\omega\, d^3k \left\{ \begin{array}{l} \alpha_e(\gamma\omega^-)[E^{sp}_{z,\omega\mathbf{k}} + \beta B^{sp}_{y,\omega\mathbf{k}}] + \\ +\beta\alpha_m(\gamma\omega^-)[B^{sp}_{y,\omega\mathbf{k}} + \beta E^{sp}_{z,\omega\mathbf{k}}] \end{array} \right\} \cdot \exp(-i\omega^- t) \tag{C3}$$

$$m_x^{ind}(t) = \gamma^{-1}(2\pi)^{-4} \iint d\omega\, d^3k\, \alpha_m(\gamma\omega^-) B^{sp}_{x,\omega\mathbf{k}} \cdot \exp(-i\omega^- t) \tag{C4}$$

$$m_y^{ind}(t) = \gamma(2\pi)^{-4} \iint d\omega\, d^3k \left\{ \begin{array}{l} \alpha_m(\gamma\omega^-)[B^{sp}_{y,\omega\mathbf{k}} + \beta E^{sp}_{z,\omega\mathbf{k}}] + \\ +\beta\alpha_e(\gamma\omega^-)[E^{sp}_{z,\omega\mathbf{k}} + \beta B^{sp}_{y,\omega\mathbf{k}}] \end{array} \right\} \cdot \exp(-i\omega^- t) \tag{C5}$$

$$m_z^{ind}(t) = \gamma(2\pi)^{-4} \iint d\omega\, d^3k \left\{ \begin{array}{l} \alpha_m(\gamma\omega^-)[B^{sp}_{z,\omega\mathbf{k}} - \beta E^{sp}_{y,\omega\mathbf{k}}] - \\ -\beta\alpha_e(\gamma\omega^-)[E^{sp}_{y,\omega\mathbf{k}} - \beta B^{sp}_{z,\omega\mathbf{k}}] \end{array} \right\} \cdot \exp(-i\omega^- t) \tag{C6}$$

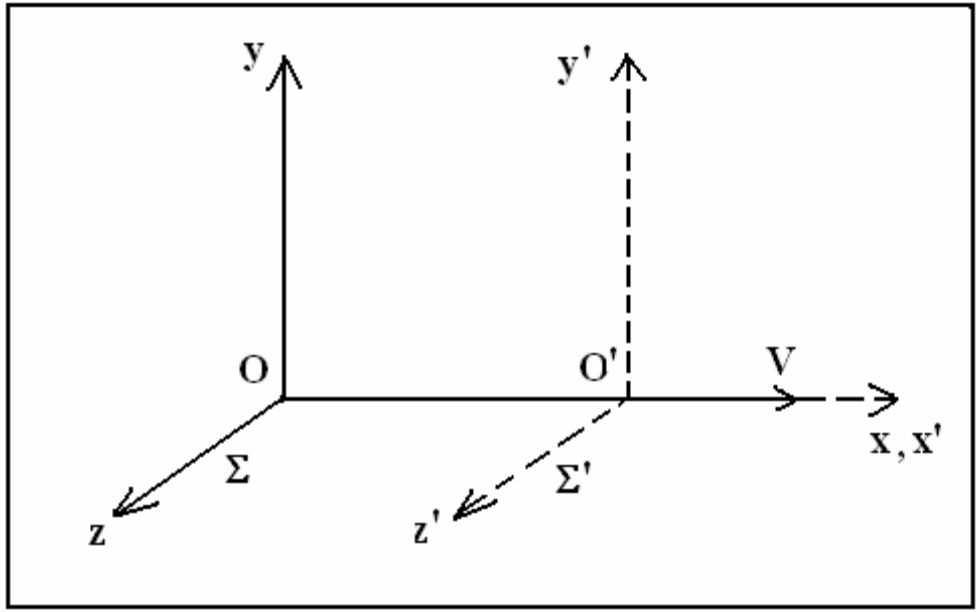

Fig.1 Cartesian reference frame $\Sigma$ associated with resting vacuum background and the particle rest frame $\Sigma'$.